\documentstyle[12pt]{article}
\newlength{\extraspace}
\newlength{\extraspaces}
\newcommand{\be}{\begin{equation}
\addtolength{\abovedisplayskip}{\extraspaces}
\addtolength{\belowdisplayskip}{\extraspaces}
\addtolength{\abovedisplayshortskip}{\extraspace}
\addtolength{\belowdisplayshortskip}{\extraspace}}
\newcommand{\ee}{\end{equation}}
\newcommand{\ba}{\begin{eqnarray}
\addtolength{\abovedisplayskip}{\extraspaces}
\addtolength{\belowdisplayskip}{\extraspaces}
\addtolength{\abovedisplayshortskip}{\extraspace}
\addtolength{\belowdisplayshortskip}{\extraspace}}
\newcommand{\ea}{\end{eqnarray}}
\newcommand{\nonu}{\nonumber \\[.5mm]}
\newcommand{\A}{&\!\!\!}
\begin{document}
\thispagestyle{empty}
\setlength{\baselineskip}{6mm}
\begin{flushright}
SIT-LP-16/07 \\
July, 2016
\end{flushright}
\vspace{7mm}
\begin{center}
{\large\bf On the role of the commutator algebra \\[2mm]
for nonlinear supersymmetry
} \\[20mm]
{\sc Kazunari Shima}
\footnote{
\tt e-mail: shima@sit.ac.jp} \ 
and \ 
{\sc Motomu Tsuda}
\footnote{
\tt e-mail: tsuda@sit.ac.jp} 
\\[5mm]
{\it Laboratory of Physics, 
Saitama Institute of Technology \\
Fukaya, Saitama 369-0293, Japan} \\[20mm]
\begin{abstract}
We discuss a closure of commutator algebras for general functionals 
in terms of Nambu-Goldstone fermions and their derivative terms 
under nonlinear supersymmetry (NLSUSY) both in flat spacetime and in curved spacetime. 
We point out that variations of functionals for vector supermultiplets 
(uniquely) determine general LSUSY transformations for linear vector supermutiplets 
with general auxiliary fields in extended SUSY, 
where the closure of the commutator algebras for NLSUSY plays a crucial role. 
\\[5mm]
%
%
\end{abstract}
\end{center}

\newpage

\noindent
Supersymmetry (SUSY) is realized linearly and nonlinearly. 
Linear supersymmetric (LSUSY) theories in flat spacetime \cite{WZ,WB} are given from the scalar and vector supermultiplets and their actions, 
while the Volkov-Akulov (VA) nonlinear supersymmtric (NLSUSY) theory \cite{VA} is expressed only in terms of Nambu-Goldstone (NG) fermions. 

The LSUSY gives plain physical pictures but the meaning of the SUSY breaking is unclear. 
On the other hand, the NLSUSY gives less physical pictures but gives the robust SUSY breaking. 
The LSUSY and NLSUSY theories are related to each other 
and their relations are shown expicitly in $N = 1$ and $N = 2$ SUSY models 
by means of the linearization of NLSUSY \cite{IK}-\cite{STT2}. 
In the relation between linear and nonlinear realizations of SUSY (NL/LSUSY relation) in flat spacetime, 
component fields of supermultiplets are expressed as functionals (composites) in terms of the NG fermions, 
which reproduce LSUSY transformations in LSUSY multiplets 
from the variations of those functionals under NLSUSY transformations for the NG fermions, 
and LSUSY actions are related to a NLSUSY one. 

On the other hand, a (global) NLSUSY transformations for the NG fermions with a vierbein field 
is generalized to curved spacetime and an Einstein-Hilbert-type NLSUSY invariant action 
are constructed in NLSUSY general relativity (GR) \cite{KS,ST1}. 
The NLSUSY-GR action containes the VA NLSUSY action in the cosmological term 
and its implications for the low energy physics is extracted from the NLSUSY action for $N \ge 2$ SUSY. 
Therefore, in order to discuss the physical consequences of NLSUSY GR, 
it is an important and interesting problem to know through the NL/LSUSY relation 
the general structure of linear supermultiplets in extended SUSY 
which includes {\it general features of auxiliary fields} prior to adopting 
a specific gauge {\it a la} the Wess-Zumino gauge. 

Towards understanding the general NL/LSUSY relation for $N \ge 2$ SUSY theories 
and obtaining a general LSUSY supermultiplet which contributes to the linearized action of NLSUSY-GR theory, 
we discuss in this letter commutator algebras for general functionals in terms of the NG fermions 
under the NLSUSY transformations both in flat spacetime and in curved spacetime. 
In flat spacetime, we explicitly show that the commutator algebra for all Lorentz-tensor functionals 
of NG fermions and their first- and higher-order derivatives are closed. 
In curved spacetime, we also show a closure of the commutator algebra 
for all Lorentz- and Riemann-tensor functinals of the NG fermions, the veirbein and their first-order derivatives. 
Based on these arguments, we point out that variations of functionals for vector supermultiplets 
(uniquely) determine general LSUSY transformations for linear vector supermultiplets in extended SUSY 
containing general auxiliary fields. 

Let us briefly review the VA NLSUSY model \cite{VA} in extended SUSY. 
The fundamental action in terms of (Majorana) NG fermions $\psi^i$ 
\footnote{
The index $i,j,\cdots$ runs from $1$ to $N$ 
and Minkowski spacetime indices are denoted by $a, b, \cdots = 0, 1, 2, 3$. 
The Minkowski spacetime metric is $\eta^{ab} = {\rm diag}(+,-,-,-) = {1 \over 2} \{ \gamma^a, \gamma^b \}$. 
Riemann spacetime indices $\mu, \nu, \cdots = 0, 1, 2, 3$ are used in arguments for curved spacetime. 
}
is written as 
\be
S_{\rm NLSUSY} = - {1 \over {2 \kappa^2}} \int d^4x \ \vert w \vert, 
\label{NLSUSYaction}
\ee
where $\kappa$ is a dimensional constant whose dimension is (mass)$^{-2}$ 
and the determinant $\vert w \vert$ is defined as 
\be
\vert w \vert = \det(w^a{}_b) = \det(\delta^a_b + t^a{}_b) 
\ee
with $t^a{}_b = - i \kappa^2 \bar\psi^i \gamma^a \partial_b \psi^i$. 
The $N$ NLSUSY action (\ref{NLSUSYaction}) is invariant under NLSUSY transformations of $\psi^i$, 
\be
\delta_\zeta \psi^i = {1 \over \kappa} \zeta^i 
- i \kappa \bar\zeta^j \gamma^a \psi^j \partial_a \psi^i, 
\label{NLSUSY}
\ee
which are parametrized by means of constant (Majorana) spinor parameters $\zeta^i$. 
The NLSUSY transformations (\ref{NLSUSY}) satisfy a closed commutator algebra, 
\be
[\delta_{\zeta_1}, \delta_{\zeta_2}] = \delta_P(\Xi^a), 
\label{NLSUSYcomm}
\ee
where $\delta_P(\Xi^a)$ means a translation with the parameters 
$\Xi^a = 2 i \bar\zeta_1^i \gamma^a \zeta_2^i$. 

Here we consider Lorentz-tensor functionals in terms of $\psi^i$ and its first- 
and higher-order derivatives ($\partial \psi^i$, $\partial^2 \psi^i$, $\cdots$, $\partial^n \psi^i$), 
\be
f^I_A = f^I_A(\psi^i, \partial \psi^i, \partial^2 \psi^i, \cdots, \partial^n \psi^i), 
\ \ \ g^J_B = g^J_B(\psi^i, \partial \psi^i, \partial^2 \psi^i, \cdots, \partial^n \psi^i) 
\label{functionals0}
\ee
where the Lorentz indices $A,B$ mean $a, ab, \cdots, {\rm etc.}$ 
and the internal indices $I,J$ are $i, ij, \cdots, {\rm etc.}$ 
If we assume that they satsify the commutator algebra (\ref{NLSUSYcomm}) as 
\be
[\delta_{\zeta_1}, \delta_{\zeta_2}] f^I_A = \Xi^a \partial_a f^I_A, 
\ \ \ [\delta_{\zeta_1}, \delta_{\zeta_2}] g^J_B = \Xi^a \partial_a g^J_B. 
\label{NLSUSYcomm2}
\ee
then, we can show that the commutator algebra for their products $f^I_A g^J_B$ are also closed as 
\be
[\delta_{\zeta_1}, \delta_{\zeta_2}] (f^I_A g^J_B) = \Xi^a \partial_a (f^I_A g^J_B). 
\ee
This means that the commutator algebra for all Lorentz-tensor functionals 
expressed by means of Eq.(\ref{functionals0}) are closed same as Eq.(\ref{NLSUSYcomm}), 
since each field of ($\psi^i$, $\partial \psi^i$, $\partial^2 \psi^i$, $\cdots$, $\partial^n \psi^i$) 
satisfies the algebra (\ref{NLSUSYcomm}), respectively. 

Based on these discussions for the closure of the commutator algebra 
under the NLSUSY transformations (\ref{NLSUSY}), 
let us examine the variations of functionals $f^I_A(\psi^i, \vert w \vert)$ which correspond to 
those of a $N = 2$ vector supermultiplet in $d = 2$ \cite{ST2,ST3}. 
In order to realize the commutator algebra (\ref{NLSUSYcomm}), 
the two supertransformations of Eq.(\ref{functionals0}) have a general form, 
\be
\delta_{\zeta_1} \delta_{\zeta_2} f^I_A = {1 \over 2} \Xi^a \partial_a f^I_A 
+ \bar\zeta_1^k \gamma^B \zeta_2^l \partial^2 h^{Ikl}_{AB}, 
\label{second-f}
\ee
where $h^{Ikl}_{AB} = h^{Ikl}_{AB}(\psi^i, \vert w \vert)$ are functionals 
factorized with respect to $\bar\psi^k \gamma_B \psi^l$. 
Since the second terms $\bar\zeta_1^k \gamma^B \zeta_2^l \ h^{Ikl}_{AB}$ are symmetric 
for exchanging the indices 1 and 2, they vanish in the commulator algebra (\ref{NLSUSYcomm2}), respectively. 
On the other hand, let us write the variations of $f^I_A$ in a simple form as 
\be
\delta_\zeta f^I_A = i \!\!\not\!\partial b^{Ik}_{AC} \gamma^C \zeta^k, 
\label{first-f}
\ee
by using functionals $b^{Ik}_{AC} = b^{Ik}_{AC}(\psi^i, \vert w \vert)$. 
In Eq.(\ref{first-f}), when the $f^I_A$ are fermionic functionals, the $b^{Ik}_{AC}$ are bosonic ones and vice versa. 
Then, comparing Eq.(\ref{second-f}) and Eq.(\ref{first-f}) determines general LSUSY transformations of $b^{Ik}_{AC}$ 
by regarding ($f^I_A$, $h^{Ikl}_{AB}$, $b^{Ik}_{AC}$) as fermionic or bosonic components, e.g. as follows; 
\be
\delta_\zeta b^{Ik}_{AC} = \gamma_C f^I_A \zeta^k 
+ i \!\!\not\!\partial h^{Ikl}_{AB} \gamma_C{}^B \zeta^l. 
\label{first-b}
\ee
Therefore, we point out through the above heuristic arguments in extended SUSY 
that general LSUSY transformations for linear supermultiplets with general auxiliary fields 
are determined by means of the functionals $f^I_A$, $h^{Ikl}_{AB}$ and $b^{Ik}_{AC}$ 
as a result of the closure of commutator algebra (\ref{NLSUSYcomm}). 

As a simple example of the above commutator-based LSUSY transformations of the composite fields, 
we can demonstrate the NL/LSUSY relation for the $d = 2$, $N = 2$ vector supermultiplet \cite{ST2,ST3}. 

The $N = 2$ LSUSY invariant (free) action with a $D$ term is written as 
\be
S_{N=2\ {\rm gauge}} = \int d^2x \left\{ -{1 \over 4} (F_{0ab})^2 + {i \over 2} \bar\lambda_0^i \!\!\not\!\partial \lambda_0^i 
+ {1 \over 2} (\partial_a A_0)^2 + {1 \over 2} (\partial_a \phi_0)^2 + {1 \over 2} D_0^2 
- {\xi \over \kappa} D_0 \right\}, 
\label{action}
\ee
where $(v_{0a}, \lambda_0^i, A_0, \phi_0, D_0)$ are components of the $N = 2$ minimal (Wess-Zumino) gauge supermultiplet 
defined by using auxiliary fields $(D, \Lambda^i, C)$ and the recombination of the components of the larger vector supermultiplet 
as follows; 
\be
(v_{0a}, \lambda_0^i, A_0, \phi_0, D_0) 
= (v_a, \lambda^i + i \!\!\not\!\partial \Lambda^i, M^{ii}, \phi, D + \Box C). 
\label{fields}
\ee
The action (\ref{action}) is invariant under LSUSY transformations for the most general components 
of the $N = 2$ vector supermultiplet in Eq.(\ref{fields}), 
which satify the algebra (\ref{NLSUSYcomm}), as follows; 
\ba
\A \A 
\delta_\zeta \lambda^i 
= D \zeta^i - i \!\!\not\!\partial M^{ij} \zeta^j 
- {i \over 2} \epsilon^{ij} \gamma_5 \!\!\not\!\partial \phi \zeta^j 
- {1 \over 2} \epsilon^{ij} \gamma^a \!\!\not\!\partial v_a \zeta^j, 
\nonu
\A \A 
\delta_\zeta D = -i \bar\zeta^i \!\!\not\!\partial \lambda^i, 
\nonu
\A \A 
\delta_\zeta M^{11} = \bar\zeta^1 \lambda^1 + i \bar\zeta^2 \!\!\not\!\partial \Lambda^2, 
\nonu
\A \A 
\delta_\zeta M^{22} = \bar\zeta^2 \lambda^2 + i \bar\zeta^1 \!\!\not\!\partial \Lambda^1, 
\nonu
\A \A 
\delta_\zeta \phi = -\epsilon^{ij} (\bar\zeta^i \gamma_5 \lambda^j 
+ i \bar\zeta^i \gamma_5 \!\!\not\!\partial \Lambda^j), 
\nonu
\A \A 
\delta_\zeta v_a
= -\epsilon^{ij} (i \bar\zeta^i \gamma_a \lambda^j 
+ \bar\zeta^i \!\!\not\!\partial \gamma_a \Lambda^j), 
\nonu
\A \A 
\delta_\zeta \Lambda^i
= M^{ij} \bar\zeta^j - M^{jj} \bar\zeta^i 
+ {1 \over 2} \epsilon^{ij} \phi \gamma_5 \zeta^j 
- {i \over 2} \epsilon^{ij} v_a \gamma^a \zeta^j 
- i \!\!\not\!\partial C \zeta^i, 
\nonu
\A \A 
\delta_\zeta C = \bar\zeta^i \Lambda^i. 
\label{LSUSY}
\ea
where a general auxiliary field $M^{12}$ appears in $\delta_\zeta \lambda^i$ and $\delta_\zeta \Lambda^i$ 
and its LSUSY transformation is given as 
\be
\delta_\zeta M^{12} = \bar\zeta^{(1} \lambda^{2)} - i \bar\zeta^{(1} \!\!\not\!\partial \Lambda^{2)}. 
\label{LSUSY2}
\ee
By means of the linearization of $N = 2$ NLSUSY in $d = 2$ \cite{ST2}, 
the LSUSY transformations (\ref{LSUSY}) and (\ref{LSUSY2}) are obtained 
from the following composite expressions of the basic fields (except for constant shifts), 
\ba
\A \A 
\lambda^i(\psi) = \xi \psi^i \vert w \vert, 
\nonu
\A \A 
D(\psi) = {\xi \over \kappa} \vert w \vert, 
\nonu
\A \A 
M^{ij}(\psi) = {1 \over 2} \xi \kappa \bar\psi^i \psi^j \vert w \vert, 
\nonu
\A \A 
\phi(\psi) = -{1 \over 2} \xi \kappa \epsilon^{ij} \bar\psi^i \gamma_5 \psi^j \vert w \vert, 
\nonu
\A \A 
v_a(\psi) = -{i \over 2} \xi \kappa \epsilon^{ij} \bar\psi^i \gamma_a \psi^j \vert w \vert. 
\nonu
\A \A 
\Lambda^i(\psi) = -{1 \over 2} \xi \kappa^2 \psi^i \bar\psi^j \psi^j \vert w \vert, 
\nonu
\A \A 
C(\psi) = -{1 \over 8} \xi \kappa^3 \bar\psi^i \psi^i \bar\psi^j \psi^j \vert w \vert, 
\label{composites}
\ea
under the $N = 2$ NLSUSY transformations. 
We have also shown \cite{ST2} the relation between the LSUSY action of Eq.(\ref{action}) 
and the NLSUSY one of Eq.(\ref{NLSUSYaction}) for $N = 2$ SUSY in $d = 2$, 
namely, the relation $S_{N=2\ {\rm gauge}} + [{\rm surface\ term}] = \xi^2 S_{N=2\ {\rm NLSUSY}}$. 

As for the above NL/LSUSY relation, we have found \cite{ST4} that 
an ansatz $\lambda^i(\psi) = \psi^i \vert w \vert$ in extended SUSY, 
which is the leading order of supercharges $Q^i$ of SUSY, 
gives general LSUSY transformations for vector supermultiplets from functionals corresponding to Eq.(\ref{composites}), 
which are determined through Eqs.(\ref{second-f}), (\ref{first-f}) and (\ref{first-b}). 
The $N = 2$ LSUSY transformations (\ref{LSUSY}) and (\ref{LSUSY2}) 
are derived from general LSUSY transformations in extended SUSY. 
Furthermore, the commutator-based linearization from Eqs.(\ref{second-f}), (\ref{first-f}) and (\ref{first-b}) 
is practical in order to construct LSUSY supermultiplets in extended SUSY. 

In curved spacetime, an Einstein-Hilbert-type NLSUSY invariant action 
is constructed in NLSUSY GR \cite{KS,ST1} as follows; 
%
%
%
\be
{\cal S}_{\rm NLSUSYGR} = -{c^4 \over 16 \pi G} \int d^4x \ \vert w \vert (\Omega + \Lambda), 
\label{NLSUSYGRaction}
\ee
In the action (\ref{NLSUSYGRaction}), $\vert w \vert = \det w^a{}_\mu$ and $w^a{}_\mu$ is defined as a unified vierbein 
$w^a{}_\mu = e^a_\mu + t^a{}_\mu$, where $e^a_\mu$ is the vierbein in GR 
and $t^a{}_\mu = - i \kappa^2 \bar\psi^i \gamma^a \partial_\mu \psi^i$. 
The $\Omega$ means a scalar curvature in terms of ($w^a{}_\mu$, $w^\mu{}_a$) and $\Lambda$ is a cosmological constant. 
In flat spacetime ($e^a_\mu \rightarrow \delta^a_\mu$), the action (\ref{NLSUSYGRaction}) 
reduces to the VA NLSUSY action (\ref{NLSUSYaction}) 
with the dimensional constant $\kappa$ being fixed to $\displaystyle{\kappa^{-2} = {c^4 \Lambda \over 8 \pi G}}$. 

The NLSUSY-GR action (\ref{NLSUSYGRaction}) is invariant under NLSUSY transformations of $\psi^i$ and $e^a_\mu$ 
\ba
\delta_\zeta \psi^i \A = \A {1 \over \kappa} \zeta^i - i \kappa \bar\zeta^j \gamma^\mu \psi^j \partial_\mu \psi^i, 
\nonu
\delta_\zeta e^a_\mu \A = \A 2i \kappa \bar\zeta^i \gamma^\nu \psi^i \partial_{[\mu} e^a_{\nu]}, 
\label{NLSUSYGR}
\ea
which induce $GL(4,{\bf R})$ transformations of $w^a{}_\mu$, 
\be
\delta_\zeta w^a{}_\mu 
= \xi^\nu \partial_\nu w^a{}_\mu + w^a{}_\nu \partial_\mu \xi^\nu, 
\ee
with $\xi^\mu = -i \kappa \bar\zeta^i \gamma^\mu \psi^i$. 
The NLSUSY transformations (\ref{NLSUSYGR}) satisfy the following commutator algebra, 
\be
[\delta_{\zeta_1}, \delta_{\zeta_2}] = \delta_{GL(4,{\bf R})}(\Xi^\mu), 
\label{NLSUSYGRcomm}
\ee
where $\delta_{GL(4,{\bf R})}(\Xi^\mu)$ is a $GL(4,{\bf R})$ transformation with parameters 
\be
\Xi^\mu = 2(i \bar\zeta_1^i \gamma^\mu \zeta_2^i - \xi_1^\nu \xi_2^\rho e_a^\mu \partial_{[\nu} e^a_{\rho]}). 
\ee
The NLSUSY-GR action (\ref{NLSUSYGRaction}) posesses large symmetries accomodating $SO(N)$ SP group, 
as follows; 
\ba
\A \A 
[{\rm global \ NLSUSY}] \otimes [{\rm local}\ GL(4,{\bf R})] \otimes [{\rm local \ Lorentz}] 
\nonu
\A \A 
\otimes [{\rm local \ spinor \ translation}] 
\otimes [{\rm global}\ SO(N)] \otimes [{\rm local}\ U(1)^N] \otimes [{\rm chiral}]. 
\ea

Similar to the argument in flat spacetime, we discuss below possible functionals of $\psi^i$ and $e^a{}_\mu$ 
under the commutator algebra (\ref{NLSUSYGRcomm}). 
For simplicity, let us first consider Lorentz- and Riemann-tensor functionals of $\psi^i$ and $e^a{}_\mu$ 
without their derivative terms, 
\be
f^{IA}{}_M = f^{IA}{}_M(\psi^i, e^a{}_\mu), 
\ \ \ g^{JB}{}_N = g^{JB}{}_N(\psi^i, e^a{}_\mu), 
\label{functionals}
\ee
with Riemann spacetime indices $M, N = \mu, \mu \nu, \cdots$, 
which satisfy the NLSUSY algebra (\ref{NLSUSYGRcomm}), i.e. 
\ba
\A \A 
[\delta_{\zeta_1}, \delta_{\zeta_2}] f^{IA}{}_M = \Xi^\rho \partial_\rho f^{IA}{}_M 
+ \sum_k f^{IA}_{M_k(\rho)} \partial_{\mu_k} \Xi^\rho, 
\nonu
\A \A 
[\delta_{\zeta_1}, \delta_{\zeta_2}] g^{JB}{}_N = \Xi^\rho \partial_\rho g^{JB}{}_N 
+ \sum_k g^{JB}_{N_k(\rho)} \partial_{\mu_k} \Xi^\rho. 
\label{algebra}
\ea
In Eq.(\ref{algebra}), Riemann spacetime indices $M_k(\rho)$ are defiend as 
$M_k(\rho) = \mu_1 \mu_2 \cdots \rho \cdots \mu_n$ and the summation $\displaystyle{\sum_k}$ means 
\ba
\sum_k f^{IA}_{M_k(\rho)} \partial_{\mu_k} \Xi^\rho 
\A = \A f^{IA}_{M_1(\rho)} \partial_{\mu_1} \Xi^\rho + f^{IA}_{M_2(\rho)} \partial_{\mu_2} \Xi^\rho + \cdots 
\nonu
\A = \A f^{IA}_{\rho \mu_2 \cdots} \partial_{\mu_1} \Xi^\rho + f^{IA}_{\mu_1 \rho \cdots} \partial_{\mu_2} \Xi^\rho + \cdots. 
\ea
Then, the commutator algebra for the products $f^{IA}{}_M g^{JB}{}_N$ are also closed as 
\ba
[\delta_{\zeta_1}, \delta_{\zeta_2}] (f^{IA}{}_M g^{JB}{}_N) 
\A = \A \Xi^\rho \partial_\rho (f^{IA}{}_M g^{JB}{}_N) 
+ \sum_k f^{IA}_{M_k(\rho)} \partial_{\mu_k} \Xi^\rho g^{JB}{}_N 
+ f^{IA}{}_M \sum_k g^{JB}_{N_k(\rho)} \partial_{\mu_k} \Xi^\rho 
\nonu
\A = \A \delta_{GL(4,{\bf R})} (f^{IA}{}_M g^{JB}{}_N)(\Xi^\rho). 
\label{algebra2}
\ea
Since the NLSUSY transformations (\ref{NLSUSYGR}) satisfy the algebra (\ref{NLSUSYGRcomm}), 
all Lorentz- and Riemann-tensor functionals expressed by means of (\ref{functionals}) satisfy the algebra (\ref{NLSUSYGRcomm}). 

Next we consider functionals including derivatives of $\psi^i$ and $e^a{}_\mu$. 
As for commutator algebras for their derivative terms under the NLSUSY transformations (\ref{NLSUSYGR}), 
only for $\partial_\mu \psi^i$ and $\partial_{[\mu} e^a{}_{\nu]}$ are closed as 
\ba
[\delta_{\zeta_1}, \delta_{\zeta_2}] \partial_\mu \psi^i 
\A = \A \Xi^\nu \partial_\nu (\partial_\mu \psi^i) + (\partial_\nu \psi^i) \partial_\mu \Xi^\nu
= \delta_{GL(4,{\bf R})} (\partial_\mu \psi^i)(\Xi^\nu), 
\nonumber\\[2mm]
[\delta_{\zeta_1}, \delta_{\zeta_2}] \partial_{[\mu} e^a{}_{\nu]} 
\A = \A \Xi^\rho \partial_\rho (\partial_{[\mu} e^a{}_{\nu]}) 
+ (\partial_{[\rho} e^a{}_{\nu]}) \partial_\mu \Xi^\rho 
+ (\partial_{[\mu} e^a{}_{\rho]}) \partial_\nu \Xi^\rho 
\nonu
\A = \A \delta_{GL(4,{\bf R})} (\partial_{[\mu} e^a{}_{\nu]})(\Xi^\nu). 
\label{algebra3}
\ea
The commutator algebra for the higher-order derivative terms of $\psi^i$ and $e^a{}_\mu$ 
(i.e. for ($\partial^2 \psi^i$, $\partial^2 e^a{}_\mu$, $\cdots$)) are not closed. 
Therefore, in curved spacetime the following Lorentz- and Riemann-tensor functionals 
expressed only in terms of $\psi^i$, $e^a{}_\mu$ and their first-order derivatives, 
\be
f^{IA}{}_M = f^{IA}{}_M(\psi^i, e^a{}_\mu; \partial_\mu \psi^i, \partial_{[\mu} e^a{}_{\nu]}) 
\label{functionals2}
\ee
satisfy the commutator algeba (\ref{NLSUSYGRcomm}) 
from the same arguments as Eqs.(\ref{algebra}) and (\ref{algebra2}). 

Furthermore, in the same way as the commutator-based arguments in flat spacetime, 
we expect that the two supertransformations for some functionals of $\psi^i$ and $e^a_\mu$ 
determine general (global) LSUSY transformations for linear supermultiplets in curved spacetime, 
which are caused by the closure of the commurtator algebra (\ref{NLSUSYGRcomm}). 

We summarize our results as follows. 
In this letter, under the NLSUSY transformations (\ref{NLSUSY}) and (\ref{NLSUSYGR}) 
we have shown the closure of the commutator algebra 
for the Lorentz-tensor functionals (\ref{functionals0}) in flat spacetime 
and for the Lorentz- and Riemann-tensor functionals (\ref{functionals2}) in curved spacetime. 
Based on those arguments, we have pointed out that the commutator-based LSUSY transformations (\ref{first-b}) 
are (uniquely) determined from the two supertransformations (\ref{second-f}) and the variations (\ref{first-f}), 
which lead to linear supermultiplets with general auxiliary fields. 
The general LSUSY supermultiplets as shown in the example of Eqs.(\ref{LSUSY}), (\ref{LSUSY2}) and (\ref{composites}) 
are constructed from examining the commutator algebra (\ref{NLSUSYcomm2}) for NLSUSY 
(and also Eq.(\ref{NLSUSYGRcomm}) for NLSUSY GR) prior to transforming to gauge supermultiplets. 
In particular, the commutator-based linearization from Eqs.(\ref{second-f}), (\ref{first-f}) and (\ref{first-b}) 
is practical in order to find general LSUSY supermultiplets in extended SUSY.

\newpage

%
\newcommand{\NP}[1]{{\it Nucl.\ Phys.\ }{\bf #1}}
\newcommand{\PL}[1]{{\it Phys.\ Lett.\ }{\bf #1}}
\newcommand{\CMP}[1]{{\it Commun.\ Math.\ Phys.\ }{\bf #1}}
\newcommand{\MPL}[1]{{\it Mod.\ Phys.\ Lett.\ }{\bf #1}}
\newcommand{\IJMP}[1]{{\it Int.\ J. Mod.\ Phys.\ }{\bf #1}}
\newcommand{\PR}[1]{{\it Phys.\ Rev.\ }{\bf #1}}
\newcommand{\PRL}[1]{{\it Phys.\ Rev.\ Lett.\ }{\bf #1}}
\newcommand{\PTP}[1]{{\it Prog.\ Theor.\ Phys.\ }{\bf #1}}
\newcommand{\PTPS}[1]{{\it Prog.\ Theor.\ Phys.\ Suppl.\ }{\bf #1}}
\newcommand{\AP}[1]{{\it Ann.\ Phys.\ }{\bf #1}}

\end{document}